\documentclass[prd,twocolumn,showpacs,nofootinbib, aps]{revtex4-1}
\usepackage{bm}
\usepackage{graphicx}
\usepackage{multirow}
\usepackage{amsmath}
\usepackage{amssymb}
\usepackage{tabulary}
\usepackage{hyperref}
\usepackage{color}
\usepackage{xspace}
\usepackage{verbatim}
\usepackage{mathtools}
\usepackage{booktabs}
\usepackage[utf8]{inputenc}

\newcommand{\mres}{m_\mathrm{B}}
\newcommand{\mgg}{m_{\gamma \gamma}}

\begin{document}

\title{Re-examining the significance of the 750 GeV diphoton excess at ATLAS}
\author{Bradley J. Kavanagh}
\email{bradley.kavanagh@lpthe.jussieu.fr}
\affiliation{LPTHE, CNRS, UMR 7589, 4 Place Jussieu, F-75252, Paris, France}
\affiliation{Institut de physique th\'eorique, Universit\'e Paris Saclay, CNRS, CEA, F-91191 Gif-sur-Yvette, France}

\begin{abstract}
The excess seen in the diphoton channel at around 750 GeV by both ATLAS and CMS has caused a great deal of excitement in the particle physics community. However, there has recently been much discussion about uncertainties in the significance of the peak seen by the ATLAS experiment. In this note, we aim to estimate this significance using a range of possible parametrisations for the smooth diphoton background. We obtain a local significance close to that reported by ATLAS and further demonstrate that the significance of the excess is not substantially reduced when more complicated background functions are considered. In particular, the background contribution is strongly constrained by the small numbers of events at large diphoton invariant mass. Future data releases will improve constraints on the diphoton background, as well as clarifying the true nature of the 750 GeV excess.
\end{abstract}

\maketitle

\section{Introduction}
\label{sec:introduction}

In the first results from Run-II of the LHC, both the ATLAS and CMS collaborations reported an excess above the expected background in the diphoton channel \cite{ATLAS1, CMS1} at a diphoton invariant mass of $\mgg \sim 750 \text{ GeV}$. The local significance of this excess was reported as $3.9\sigma$ ($2.3\sigma$ global) by ATLAS and $2.6\sigma$ ($1.2\sigma$ global) by CMS. The possibility that this excess could be due to a new bosonic particle has caused a great deal of excitement in the particle physics community and the list of papers exploring the implications for New Physics is of course too long to mention.

Since the announcement of these results \cite{Run2Seminar}, a number of authors have performed fits to the Run-II diphoton data from a theory perspective in order to estimate the mass, width and cross section times branching ratio of a possible new particle (see e.g.~Ref.~\cite{Falkowski:2015swt}). However, these analyses typically assume a fixed form for the diphoton background whose parameters are not included in the fits. This procedure is not strictly correct, as parameter values extracted from the data should take into account uncertainties in the background. A full likelihood analysis of both the ATLAS and CMS binned diphoton data was performed in Ref.~\cite{Buckley:2016mbr}, assuming the same background models used by the respective collaborations. The local significances reported therein are in broad agreement with those reported by ATLAS and CMS.

However, a recent paper by Davis et al.~\cite{Davis:2016hlw} explored the possibility of allowing more freedom in the modelling of the ATLAS continuum diphoton background. Their conclusion was that the background could be fit by a number of different functional forms and that some of these forms lead to a lower significance for the diphoton excess than that reported by ATLAS, perhaps as low as $2.0\sigma$. Motivated by this discrepancy between the results of the ATLAS collaboration and those of Davis et al., we perform an independent analysis of the binned ATLAS diphoton invariant mass spectrum to estimate the significance of the excess under different background assumptions. In particular, we perform a Poissonian likelihood fit over 40 bins in the diphoton invariant mass $\mgg$, making this analysis distinct from that of Davis et al. The code used to perform this analysis is publicly available online \cite{Website}.

The intention of this note is not to improve upon the ATLAS analysis or to report definitive values for the significance of the 750 GeV excess. Indeed, this is not possible with the information which is publicly available. Instead, we aim to explore the robustness of the excess to variations in background fitting, as well as to clarify certain aspects of the fits performed by ATLAS. 

The different functional forms we consider for the diphoton background are given in Eq.~\ref{eq:functions}, while the local significances we obtain for the 750 GeV diphoton excess (using each functional form) are given in Table~\ref{tab:significance}. We describe in detail the fitting procedure used in this work in Sec.~\ref{sec:fitting}, followed by the results of background-only and signal + background fits in Sec.~\ref{sec:BG-only} and Sec.~\ref{sec:Sig+BG}. Finally, in Sec.~\ref{sec:Discussion} we discuss the implications of these results for the significance of the 750 GeV excess.

\section{Fitting procedure}
\label{sec:fitting}

Within a frequentist framework, the significance of a possible excess in the diphoton spectrum can be estimated by calculating the maximum likelihood obtained assuming that only backgrounds contribute to the observed events (background-only) and that obtained assuming that both signal and background have a contribution to the observed event rate (signal + background). By comparing the maximum likelihood obtained under these two assumptions, we can then calculate the significance of a possible signal in the data.

The likelihood $\mathcal{L}(\boldsymbol{\theta})$ is simply the probability of obtaining the observed data assuming a particular set of parameters $\boldsymbol{\theta}$. For binned data, this has the Poisson form:

\begin{align}
\begin{split}
\mathcal{L}(\boldsymbol{\theta}) &= \prod_{i = 1}^{n_\mathrm{bins}} \frac{(N_e^i)^{N_o^i}}{N_o^i!} \exp(-N_e^i)\,,
\end{split}
\end{align}
where $N_e^i = N_e^i(\boldsymbol{\theta}) $ is the expected number of events in the $i^{\mathrm{th}}$ bin (for a given set of parameters under the background or signal + background hypothesis) and $N_o^i$ is the observed number of events in the $i^{\mathrm{th}}$ bin. In particular, we do not include any additional terms in the likelihood due to uncertainties on the diphoton invariant mass resolution (as in Ref.~\cite{ATLAS1}).

We use a total of $n_\mathrm{bins} = 40$ bins in the $m_{\gamma \gamma}$ invariant mass, each with width 40 GeV, spanning the range $m_{\gamma \gamma} \in [150, 1750] \, \, \mathrm{GeV}$. We obtain the number of events observed in each bin from Ref.~\cite{ATLAS1} (by digitising Fig.~1 therein using a publicly available plot digitiser \cite{WebPlot}). This procedure may introduce some digitisation error into the analysis, particularly for bins which contain a large number of events. However, we have explicitly checked the impact of digitisation errors on our results and, as described in Sec.~\ref{sec:Discussion}, these do not affect the conclusions we report.

The expected number of background events is obtained by integrating the background event distribution over each bin in $m_{\gamma \gamma}$, for a given choice of functional form and for a given set of background parameters $\boldsymbol{\theta}_b$. In Ref.~\cite{ATLAS1}, the ATLAS collaboration discuss a set of possible empirical functions for the continuum background. These have been adapted from functions used in multi-jet searches for New Physics \cite{Aad:2010ae} and have been validated against both Monte Carlo and data samples. In Ref.~\cite{Davis:2016hlw}, Davis et al.~introduced another possible parameterisation for the continuum diphoton background (also validated with a Monte Carlo study) which we discuss further in Sec.~\ref{sec:Discussion}. In this work, however, we focus on those background functions used by the ATLAS collaboration, given explicitly by

\begin{align}
\label{eq:functions}
f_k(x) &= \mathcal{N} (1-x^{1/3})^b x^{\sum_{j = 0}^k a_j (\log x)^j}
\end{align}
where $x = m_{\gamma \gamma}/\sqrt{s}$ and $f(x)$ is the differential distribution of expected background events (in units of events/40 GeV).\footnote{Note that we use the notation $\log$ for the natural logarithm $\log_e$.} We will consider $k = 0, 1, 2$ in this work, for which the parameters $b$, $a_0$, $a_1$ and $a_2$ determine the shape of the background and are allowed to vary in the analysis. The parameter $\mathcal{N}$ which controls the overall normalisation of the background is typically included when using this class of functions \cite{Aad:2010ae}. While it is not explicitly described in Ref.~\cite{ATLAS1}, the free normalisation parameter has been included in previous analyses of the diphoton channel \cite{Aad:2014eha} and its inclusion will allow us to better match the results reported by ATLAS. In light of this, we will assume that ATLAS allow for a free normalisation of the background in their analysis, though for comparison we will consider cases where $\mathcal{N}$ is set to 1 (`fixed-$\mathcal{N}$') as well as where $\mathcal{N}$ is allowed to vary (`free-$\mathcal{N}$'). 

The expected number of signal events coming from a possible new resonance near $\mgg \sim 750 \text{ GeV}$ is obtained using either the narrow width approximation (NWA) or by allowing the width of the resonance to vary freely (free-width). Under the NWA, the signal is modelled as a Gaussian centred on $\mgg = \mres$ with standard deviation $\sigma$ given by the diphoton invariant mass resolution. This resolution is taken to increase linearly from $\sigma = 2 \text{ GeV}$ at $\mgg = 200 \text{ GeV}$ to $\sigma = 13 \text{ GeV}$ at $\mgg = 2 \text{ TeV}$ \cite{ATLAS1}, giving a value of around $\sigma = 5.4 \text{ GeV}$ near the putative resonance at 750 GeV. This is a rather crude approximation to the double-sided Crystal Ball function (DCSB) used by the ATLAS collaboration in their NWA analysis. However, given that the bin widths used in this analysis are much larger than the mass resolution, we do not expect the precise shape of the narrow resonance to strongly affect our results. Indeed, we are still able to recover a significance close to that reported in Ref.~\cite{ATLAS1}. The signal is normalised so as to contribute a total of $N_S$ events, meaning that for the NWA we have two free parameters describing the signal: $\boldsymbol{\theta}_s = (\mres, N_S)$. 

In the free-width analysis, we model the signal as a Breit-Wigner distribution with width $\Gamma = \alpha \mres$. We do not include the effects of finite diphoton invariant mass resolution in the free-width analysis. However, as in the NWA analysis, we can still recover a significance close to that reported in Ref.~\cite{ATLAS1}. In the free-width analysis, we have three free parameters: $\boldsymbol{\theta}_s = (\mres, N_S, \alpha)$.

In order to explore the signal and background parameter spaces, we use the publicly available Markov Chain Monte Carlo (MCMC) software \texttt{emcee} \cite{ForemanMackey:2012ig}, which uses Affine Invariant MCMC Ensemble sampling \cite{Goodman2010}. We perform at least 20000 likelihood evaluations for each fit. The prior ranges for the parameters used in this work are listed in Table~\ref{tab:priors}, though we emphasise that \texttt{emcee} was only used to find the maximum likelihood points and therefore that the precise form of the priors should have little impact on the analysis.

\begin{table}[t]\centering
\begin{tabular}{@{}l@{\hskip 4\tabcolsep}l@{}}
\toprule
\toprule
Parameter			&	Prior range   \\
				\hline \hline
$\log_{10} \mathcal{N}$		& 	[-25, 25] \\
$b$, $a_0$, $a_1$, $a_2$ & [-25, 25]\\
$\mgg$ & [700, 800] GeV\\
$\alpha$ & [1, 10] \%\\
$N_S$ & [0.01, 100] \\
\bottomrule
\bottomrule
\end{tabular}
\caption{Prior ranges for the background and signal parameters used in this analysis. In each case, the prior is constant over the specified range. We note that because we use MCMC only to find the maximum likelihood point, the exact form of the prior does not affect the results presented.}
\label{tab:priors}
\end{table}

We quantify the significance of a possible excess by constructing the test statistic $q_0$ \cite{Cowan:2010js}:

\begin{align}
\label{eq:q0}
\begin{split}
q_0 = 
\begin{dcases}
-2\log \frac{\mathcal{L}(N_S = 0, \hat{\hat{\boldsymbol{\theta}}}_b)}{\mathcal{L}(\hat{N}_S, m_B, \alpha,\hat{\boldsymbol{\theta_b}})} & \quad \hat{N}_S \geq 0 \\
0 & \quad \hat{N}_S < 0\,,
\end{dcases}
\end{split}
\end{align}
where $\hat{\hat{\boldsymbol{\theta}}}_b$ is the set of background parameters which maximise the likelihood under the background-only hypothesis and $\hat{\boldsymbol{\theta}}$ is the set of parameters which maximise the full likelihood under the signal + background hypothesis for a given value of $m_B$ and (if we are considering the free-width analysis) $\alpha$. Here, we restrict to positive parameter values for the number of signal events, so the best fit value $\hat{N}_S$ will always be positive.

The local $p$-value for the background-only hypothesis is obtained from the observed value of the test statistic $q_{0, \mathrm{obs}}$ using,

\begin{equation}
p_0 = \int_{q_{0, \mathrm{obs}}}^\infty f(q_0) \, \mathrm{d}q_0\,
\end{equation}
where $f(q_0)$ is the probability density function of $q_0$ under the background-only hypothesis. According to Wilks' Theorem \cite{Wilks1938}, the log-likelihood ratio (appearing in the top line of Eq.~\ref{eq:q0}) is asymptotically $\chi_m^2$ distributed if the background-only hypothesis is correct. The number of degrees of freedom $m$ is given by the difference between the number of parameters in the background-only and the signal + background hypotheses. For a fixed value of $m_B$ (and $\alpha$), there is only one free parameter (the number of signal events $N_S$) so we have $m = 1$.  As detailed in Ref.~\cite{Cowan:2010js}, in this case $q_0$ follows a `half chi-square' distribution and the local signal significance is simply given by $Z = \sqrt{q_{0, \mathrm{obs}}}$. The maximum value of the local significance can then be obtained by maximising over $m_B$ (or $m_B$ and $\alpha$ in the free-width analysis).

\section{Background-only fits}
\label{sec:BG-only}

\begin{figure}[t]
\centering
\includegraphics[width=0.48\textwidth]{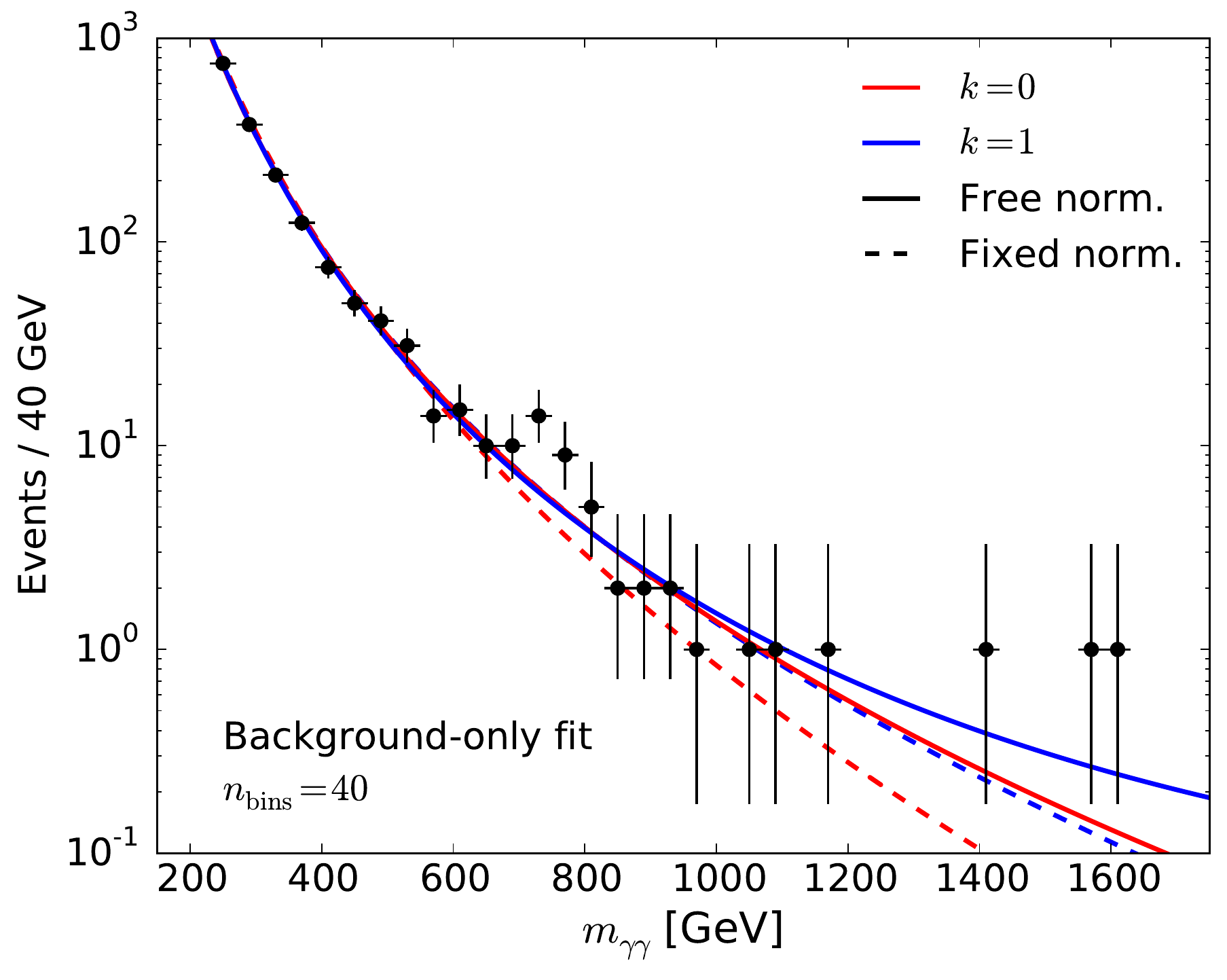}
\caption{Background-only fits to the ATLAS diphoton invariant mass spectrum. The observed numbers of events in each bin are shown as black circles, while the curves show the background distribution for the $k = 0$ (red) and $k = 1$ (blue) empirical functions defined in Eq.~\ref{eq:functions}. We show background fits with both free normalisation (solid) and with fixed normalisation (dashed). The background parameters used are those which maximise the likelihood for the background-only hypothesis. Note that the $k = 2$ background fits are not shown as they lie close to the $k = 1$, free normalisation curve (solid blue).}
\label{fig:BG-only}
\end{figure}

We begin by examining in Fig.~\ref{fig:BG-only} the fits to the ATLAS data (black points) which are obtained when no signal contribution is included. We show the best fit (maximum likelihood) background curves for $k = 0$ (red) and $k = 1$ (blue), both using a free normalisation $\mathcal{N}$ (solid lines) and when setting $\mathcal{N} = 1$ (dashed lines). We note that the error bars on the data points are for illustrative purposes and denote the $1\sigma$ confidence intervals on the mean number of expected events in each bin given the number of observed events.

As pointed out by Davis et al.~\cite{Davis:2016hlw}, the $k = 0$ fixed-normalisation background curve (dashed red) appears to underestimate the background above $\mgg > 1000 \text{ GeV}$. It is perhaps not surprising that this parametrisation is not able to give a good fit to the background over the entire invariant mass range, given that it includes only two free parameter. When adding an additional parameter to the background function, either using $k = 0$ with free normalisation (solid red) or $k = 1$ with fixed normalisation (dashed blue), the fits tend to prefer a higher background at high mass in order to alleviate this possible tension. The background contribution is also increased in the region of the 750 GeV excess, suggesting that the significance of the peak will be reduced when including these additional parameters. 

We note that both the $k = 0$, free-$\mathcal{N}$ background and the $k = 1$, fixed-$\mathcal{N}$ background appear to match closely the background-only fit reported in Fig.~1 of Ref.~\cite{ATLAS1}. The ATLAS collaboration state that they use the $k = 0$ background function in Ref.~\cite{ATLAS1} and we therefore assume that they also fit the normalisation $\mathcal{N}$ of the background, in order to match the results shown here in Fig.~\ref{fig:BG-only}. It therefore appears then that the background curve used by ATLAS fits the data well, apart from in the region of the 750 GeV excess and above $\mgg \sim 1600 \text{ GeV}$ (where ATLAS report an excess with $2.8\sigma$ local significance). 

\begin{table}[t]\centering

\begin{tabular}{@{}l@{\hskip 4\tabcolsep}l@{\hskip 4\tabcolsep}l@{\hskip 4\tabcolsep}l@{}}
\toprule
\toprule
Background function		& $N_p$		& $\log \hat{\mathcal{L}}$ & BIC \\
\hline \hline
\multicolumn{4}{l}{Fixed normalisation}\\
$\qquad k = 0$			&2	& -87.9	&  183.2 \\
$\qquad k = 1$			&3	& -82.4	& 175.9 \\
$\qquad k = 2$			&4	& -80.4     & 175.6 \\
\multicolumn{4}{l}{Free normalisation}\\
$\qquad k = 0^\dagger$			&3	& -81.9	&   174.9 \\
$\qquad k = 1$			&4	& -80.9	& 176.6 \\
$\qquad k = 2$			&5	& -80.0     & 178.4 \\
\bottomrule
\bottomrule
\end{tabular}
\caption{Number of parameters $N_p$, maximum log-likelihood $\log\hat{\mathcal{L}}$ and Bayesian Information Criterion (BIC) obtained in background-only fits to the ATLAS diphoton invariant mass spectrum using the background functions in Eq.~\ref{eq:functions}. The BIC is defined in Eq.~\ref{eq:BIC}. The background function used by the ATLAS collaboration in Ref.~\cite{ATLAS1} is marked with a dagger.}
\label{tab:BIC}
\end{table}

Following Ref.~\cite{Davis:2016hlw}, we have calculated the Bayesian Information Crierion (BIC) \cite{Schwarz1978}, a model selection criterion which be used to compare the fit to data obtained using different models, penalising models which have additional free parameters. The Akaike Information Criterion (AIC) \cite{Akaike1974} is a related model selection criterion, though in general the BIC penalises the addition of extra parameters more strongly. The BIC is defined as:

\begin{equation}
\label{eq:BIC}
\mathrm{BIC} = - 2 \log \hat{\mathcal{L}} + N_p \log n_\mathrm{bins}\,,
\end{equation}
where $\hat{\mathcal{L}}$ is the maximum likelihood and $N_p$ is the number of free parameters in the model. The lower the BIC, the stronger the evidence for the model in question. A difference of around 2 in the BIC between two different models is considered positive evidence that the model with the higher BIC should be rejected in favour of the one with the lower. A difference in BIC greater than 6 is considered strong evidence \cite{Kass1995}.

We show in Table~\ref{tab:BIC} the maximum log-likelihood and the value of the BIC obtained in background-only fits using the three parametrisations $k = 0,1,2$ with both free and fixed normalisation. We find that the $k=0$, fixed-$\mathcal{N}$ background gives the largest BIC, indicating that there is strong evidence to reject this background in favour of the alternative parametrisations. Note that the remaining BIC values are all rather close in value, indicating that there should be no strong preference amongst them. This matches the conclusion of the ATLAS collaboration \cite{ATLAS1}, which used both a Fisher test and a `spurious signal' analysis\footnote{The `spurious signal' analysis requires that (for a given functional form for the background) the bias on the fitted signal yield is significantly smaller than the statistical uncertainty on the signal yield, as determined from Monte Carlo samples. See Refs.~\cite{ATLAS1,Aad:2014eha} for further details.} to determine that no background models more complex than $k = 0$, free-$\mathcal{N}$ were necessary to fit the data.

We finally note that increasing the number of parameters further, with the $k = 2$ functional form, does not substantially improve the fits, as reflected by the slightly larger BIC value for $k=2$, free-$\mathcal{N}$. We do not show these functional forms in Fig.~\ref{fig:BG-only}, because they lie very close to the solid blue $k=1$, free-$\mathcal{N}$ curve. The smooth background at high $\mgg$ cannot be increased further due to the tension with a number of empty bins. It therefore seems difficult to further increase the background contribution to the rate at either $\mgg \sim 750 \text{ GeV}$ or $\mgg \sim 1600 \text{ GeV}$ by going to more complex smooth background parametrisations.

\section{Signal + background fits}
\label{sec:Sig+BG}

\begin{figure}[t]
\centering
\includegraphics[width=0.48\textwidth]{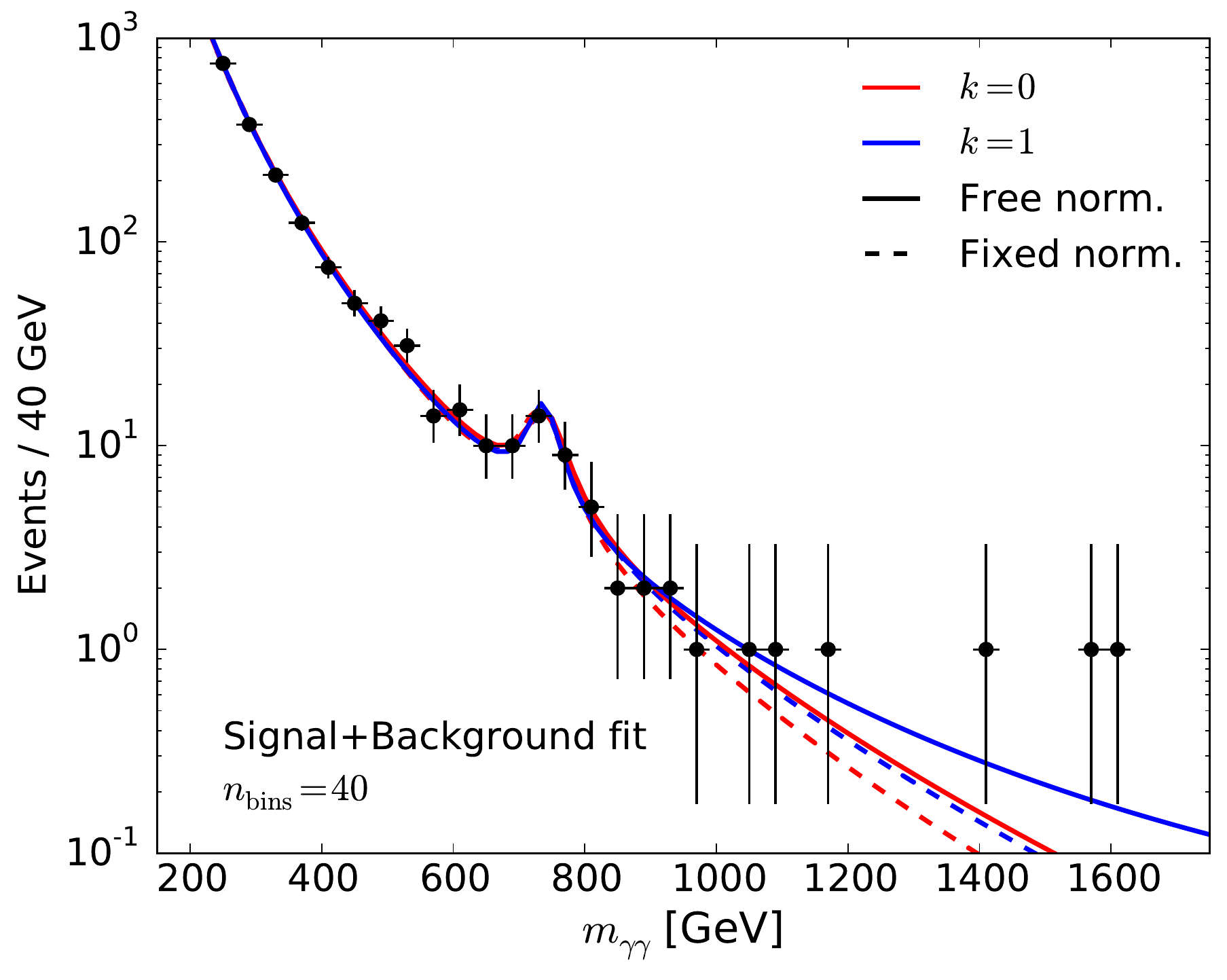}
\caption{Signal + background fits to the ATLAS diphoton invariant mass spectrum, allowing the width of the resonance near 750 GeV to vary freely. The colour scheme matches that of Fig.~\ref{fig:BG-only}. The parameters used are those which maximise the likelihood for the signal + background hypothesis. As in Fig.~\ref{fig:BG-only}, we do not show the fits for $k = 2$, as these lie close to the $k = 1$, free normalisation curve (solid blue).}
\label{fig:Sig+BG}
\end{figure}

We now examine the signal + background fits to the ATLAS diphoton spectrum and determine the significance of the diphoton excess under different assumptions for the background. In Fig.~\ref{fig:Sig+BG}, we show the maximum likelihood fits to the data for the signal + background hypothesis using the free-width analysis for the possible new resonance near 750 GeV. As before, we include fits for $k = 0, 1$ (see Eq.~\ref{eq:functions}) using both free and fixed normalisation, with the colour scheme matching that of Fig.~\ref{fig:BG-only}.

Adding a new resonance near 750 GeV clearly allows the data to be better fit with all background parametrisations. As in the background-only fits, the $k = 0$, free-$\mathcal{N}$  and $k =1$, fixed-$\mathcal{N}$ backgrounds produce similar results. In all cases, the expected rate at high $\mgg$ is reduced compared with the background-only fits. With the 750 GeV excess saturated by the signal contribution, the background above 1000 GeV can be reduced to better fit the large number of empty bins.

In Table~\ref{tab:significance}, we report the maximum local significance of the ATLAS 750 GeV excess obtained in this work, using each of the functional forms introduced in Eq.~\ref{eq:functions}. For reference, the significances reported by ATLAS in Ref.~\cite{ATLAS1} are $3.6\sigma$ (NWA) and $3.9\sigma$ (free-width). The significances obtained in this work, using the same background function used by ATLAS, are $3.4\sigma$ (NWA) and $3.6\sigma$ (free-width). We discuss in Sec.~\ref{sec:Discussion} the possible sources of this small discrepancy.

For the simplest background function ($k = 0$, fixed-$\mathcal{N}$) the significance of the excess is significantly larger.\footnote{We note that we find an even larger significance using this background than that reported by Davis et al.~\cite{Davis:2016hlw}, though given the different statistical approaches, this is perhaps not surprising.} However, in Sec.~\ref{sec:BG-only}, we found (in agreement with ATLAS) that this simple function fits the background significantly worse than the more complex ones. The remaining background parametrisations lead to very similar significances compared with the $k = 0$, free-$\mathcal{N}$ function. Thus, the significance of the excess appears to be robust against the choice of background.

\begin{table}[t]\centering
\begin{tabular}{@{}l@{\hskip 4\tabcolsep}l@{\hskip 4\tabcolsep}l@{}}
\toprule
\toprule
Background function		& NWA & Free-width \\
\hline \hline
\multicolumn{3}{l}{Fixed normalisation}\\
$\qquad k = 0$				& $4.2\sigma$	&  $4.9\sigma$ \\
$\qquad k = 1$				& $3.4\sigma$	& $3.7\sigma$ \\
$\qquad k = 2$				& $3.4\sigma$    & $3.7\sigma$ \\
\multicolumn{3}{l}{Free normalisation}\\
$\qquad k = 0^\dagger$				& $3.4\sigma$	&   $3.6\sigma$ \\
$\qquad k = 1$				& $3.5\sigma$	&   $3.8\sigma$ \\
$\qquad k = 2$		         & $3.4\sigma$    & $3.6\sigma$ \\
\hline
ATLAS reported & $3.6\sigma$ & $3.9\sigma$\\
\bottomrule
\bottomrule
\end{tabular}
\caption{Estimated local significance of the ATLAS 750 GeV diphoton excess obtained in this work using each of the background functions described in Eq.~\ref{eq:functions}, assuming a freely varying resonance width (free-width) and under the narrow width approximation (NWA). The background function used by the ATLAS collaboration in Ref.~\cite{ATLAS1} is marked with a dagger. For comparison, we also give the local significance reported by the ATLAS collaboration.}
\label{tab:significance}
\end{table}

\section{Discussion}
\label{sec:Discussion}

As initially pointed out by Davis et al.~\cite{Davis:2016hlw}, the significance of the 750 GeV diphoton excess reported by ATLAS is higher when the background is fit with the $k = 0$, fixed-normalisation parametrisation and is reduced when an extra parameter is added to the background fit. However, this appears to be consistent with the results reported by ATLAS. Assuming that the overall normalisation of the $k=0$ background is also included in the fit, we recover the background-only fit presented in Ref.~\cite{ATLAS1} and obtain significances close to (but slightly smaller than) those reported by the ATLAS collaboration.

We also note that the data show no preference for an increase in complexity of the background function (as demonstrated by the Bayesian Information Criterion study in Table~\ref{tab:BIC}). Furthermore, we find that adding additional parameters to the background fit does not have any impact on the significance of the excess. This is because of a number of bins above $\mgg \sim 1100 \text{ GeV}$ which see no events. Any smooth background is constrained not to overshoot these bins. 

Davis et al.~introduce a different possible parametrisation for the background (which was also validated by a Monte Carlo study) and find that the significance of the excess is further reduced with respect to the $k = 1$, fixed-$\mathcal{N}$ case. However, the empty bins at high $\mgg$ were not included in that analysis, leading to a background fit which overestimates the high $\mgg$ event rate. Indeed, using the Davis et al.~background parametrisation (with free normalisation) in this analysis gives a local significance of $3.8\sigma$ for a free-width resonance. This does not discount the possibility that exploring a wider range of possible background functions may impact the significance of the 750 GeV excess, but the correct constraints from the entire range of $\mgg$ should be taken into account.

It is of course necessary to point out that the significances we report are only estimates and care must be taken when comparing with the official ATLAS analysis. In particular, the results reported by ATLAS use the full unbinned data set, while we consider here only binned data. This has the largest impact on our NWA results. The diphoton invariant mass resolution ($\sim$5.4 GeV) is substantially smaller than the bin width (40 GeV), meaning that a narrow resonance cannot be resolved in the binned data. The fitted signal resonance is effectively widened by integration over a single bin, improving the fit to the relatively wide observed excess ($\sim$3 bins). This means that the significance we obtain for the NWA is larger than it would be using the unbinned data. However, the ATLAS analysis includes an additional nuisance parameter for the invariant mass resolution, which we have omitted. This mass resolution parameter can be increased to improve the fit to the data, leading to a significance similar to that obtained here.

We also emphasise that the binned data was obtained by digitising the results released in Ref.~\cite{ATLAS1}. However, we have investigated the possible impact of digitisation error on our analysis. In order to do this, we added random noise to the first 10 bins in $\mgg$ (distributed uniformly between $- 3\%$ and $+3\%$ of the number of events in each bin) in order to simulate digitisation errors.\footnote{At larger values of $\mgg$ the small number of events means that integer numbers of events can accurately be read off the figure. However at small values of $\mgg$ digitisation error could induce variations of order 10-100 events.} Ten such `randomised' datasets were generated and the peak significance for each was calculated assuming the $k = 0$, free-$\mathcal{N}$ background. The resulting local significances were in the range:

\begin{align}
\begin{split}
\text{NWA: } \quad & 3.3\sigma \mbox{--} 3.5 \sigma\\
\text{Free-width:} \quad & 3.5\sigma \mbox{--} 3.8\sigma\,.
\end{split}
\end{align}
These tests indicate that digitisation could have induced an error of order $0.2\sigma$ in the analysis, and may also explain some of the discrepancy between our results and those reported by ATLAS.

We further note that in this work we have used only approximate functional forms for the signal distributions and while we approximately recover the significances reported by the ATLAS collaboration, our analysis does not capture many of the important details involved in fitting the signal and background (for example, uncertainties in the diphoton mass reconstruction). Furthermore, we have taken an empirical approach and allowed a wide range of values for the background parameters $\mathcal{N}$, $b$, $a_0$, $a_1$, $a_2$. This may not accurately reflect the background distributions seen in MCMC and data samples. However, restricting the possible ranges of the background parameters is likely only to increase the significance of a possible excess. 

In spite of these simplifications, the broad message of this note still holds. The significance of the 750 GeV excess does not appear to be strongly affected by different choices of smooth background model. While more complicated background distributions could be explored to ensure the robustness of the significance reported by ATLAS, these are likely to be highly constrained by the lack of observed events at high diphoton invariant mass.

\acknowledgments

The author would like to thank the authors of Ref.~\cite{Davis:2016hlw} for stimulating and helpful discussions. The help of Adam Falkowski was invaluable in the completion of this work and the preparation of this manuscript. The organisers and participants of the Magic Journal Club are also thanked for interesting (if long) discussions on the 750 GeV excess. The author is supported by the John Templeton Foundation Grant 48222 and by the European Research Council ({\sc Erc}) under the EU Seventh Framework Programme (FP7/2007-2013)/{\sc Erc} Starting Grant (agreement n.\ 278234 --- `{\sc NewDark}' project). The author also acknowledges the hospitality of the Institut d'Astrophysique de Paris, where part of this work was completed.

\bibliography{Diphoton.bib}

\end{document}